\documentclass[aps,prb,twocolumn,superscriptaddress]{revtex4-1}
\usepackage{amsmath}
\usepackage[pdftex]{graphicx}
\usepackage[thinspace]{SIunits}

\providecommand \BibitemShut  [1]{\csname bibitem#1\endcsname}%

\begin{document}


\title{High fidelity optical preparation and coherent Larmor precession of a single hole in an InGaAs quantum dot molecule.}

\author{K. M\"uller}
 \affiliation{Walter Schottky Institut and Physik-Department, Technische Universit\"at M\"unchen, Am Coulombwall 4, 85748 Garching, Germany \\}
\author{A. Bechtold}
 \affiliation{Walter Schottky Institut and Physik-Department, Technische Universit\"at M\"unchen, Am Coulombwall 4, 85748 Garching, Germany \\}
\author{C. Ruppert}
 \affiliation{Experimentelle Physik 2, TU Dortmund, 44221 Dortmund, Germany \\}
\author{C. Hautmann}
 \affiliation{Experimentelle Physik 2, TU Dortmund, 44221 Dortmund, Germany \\}
\author{J. S. Wildmann}
 \affiliation{Walter Schottky Institut and Physik-Department, Technische Universit\"at M\"unchen, Am Coulombwall 4, 85748 Garching, Germany \\}
\author{T. Kaldewey}
 \affiliation{Walter Schottky Institut and Physik-Department, Technische Universit\"at M\"unchen, Am Coulombwall 4, 85748 Garching, Germany \\}
\author{M. Bichler}
 \affiliation{Walter Schottky Institut and Physik-Department, Technische Universit\"at M\"unchen, Am Coulombwall 4, 85748 Garching, Germany \\}
\author{H. J. Krenner}
 \affiliation{Lehrstuhl f\"ur Experimentalphysik 1 and Augsburg Centre for Innovative Technologies (ACIT), Universit\"at Augsburg, Universit\"atsstr. 1,
86159 Augsburg, Germany\\}
\author{G. Abstreiter}
 \affiliation{Walter Schottky Institut and Physik-Department, Technische Universit\"at M\"unchen, Am Coulombwall 4, 85748 Garching, Germany \\}
\author{M. Betz}
 \affiliation{Experimentelle Physik 2, TU Dortmund, 44221 Dortmund, Germany \\}
\author{J.J. Finley}
 \email{finley@wsi.tum.de}
 \affiliation{Walter Schottky Institut and Physik-Department, Technische Universit\"at M\"unchen, Am Coulombwall 4, 85748 Garching, Germany \\}

\date{\today}

\begin{abstract}
We employ ultrafast pump-probe spectroscopy with photocurrent readout to directly probe the dynamics of a single hole spin in a single, electrically tunable self-assembled quantum dot molecule formed by vertically stacking InGaAs quantum dots. Excitons with defined spin configurations are initialized in one of the two dots using circularly polarized picosecond pulses. The time-dependent spin configuration is probed by the spin selective optical absorption of the resulting few Fermion complex. Taking advantage of sub-5 ps electron tunneling to an orbitally excited state of the other dot, we initialize a single hole spin with a purity of $> 96 \, \%$, i.e., much higher than demonstrated in previous single dot experiments.  Measurements in a lateral magnetic field monitor the coherent Larmor precession of the single hole spin with no observable loss of spin coherence within the $\sim 300 ps$ hole lifetime. Thereby, the purity of the hole spin initialization remains unchanged for all investigated magnetic fields.

\end{abstract}


\maketitle



The spin of charge carriers trapped in semiconductor quantum dots (QDs) is highly promising for harnessing quantum phenomena for novel applications.\cite{Hanson2007} However, for electrons localized in III-V nanostructures hyperfine coupling to the fluctuating nuclear spin bath leads to rapid dephasing and, consequently, emphasis has increasingly shifted to valence-band holes since they couple $\sim 10\times$ more weakly to nuclear spins \cite{Fallahi2010,Chekhovich2011}. Moreover, for \textit{pure} heavy holes the nuclear spin coupling is of Ising-type such that dephasing can be greatly suppressed using an in-plane magnetic field\cite{Fischer2008}. Optically active QDs are particularly attractive since spin states can be directly addressed over ultrafast timescales \cite{Press2008, Ramsay08, Greilich2009, Kim2010, Atatuere2010, DeGreve2011, Greilich2011} and quantum states can be mapped to / from the optical polarization state. Vertically stacked QD-molecules with tunable interdot tunnel couplings \cite{Krenner2005, Stinaff2006, Krenner2006} allow photogenerated charges to be rapidly separated into different nanostructures\cite{Mueller2012, Gawarecki10} opening the way to exploit inelastic tunneling to photogenerate spins in the molecule to realize a high fidelity spin based quantum memory.\cite{Simon2010}

Here, we exploit ultrafast inelastic electron tunneling within a single QD-molecule to optically prepare a single hole spin and make precision readout using Pauli blockade and detecting photocurrent. An exciton with a well defined spin configuration is resonantly generated in one of the two dots forming the molecule using a circularly polarized pump beam. The polarization dynamics of the few-Fermion complex are detected using a time-delayed probe pulse that is either co- or cross-circularly polarized with respect to the pump. Our results allow us to directly interrogate both the carrier dynamics (relaxation within or tunneling out of the molecule) and the hole spin configuration over ultrafast timescales. We demonstrate fast ($<5\,ps$) high purity ($> 96\%$) initialization of a single hole spin by making use of ultrafast inelastic inter-molecular tunneling. In strong contrast to experiments with single dots, the purity of the hole spin initialization is enhanced to values above ($ 96\%$) since the e-h exchange coupling is rendered ineffective by the rapid charge separation. 
Finally, we demonstrate the coherent Larmor precession of a single hole spin in a lateral magnetic field. The in-plane hole g-factor is measured to be small ($g = 0.121\pm0.001$) and, crucially, no detectable loss of hole spin coherence is observed over the $\sim 300 ps$ hole population lifetime. These findings indicate that the hole spin coherence time extends beyond nanosecond timescales, fully consistent with  theoretical expectations\cite{Fischer2008} and the findings of refs.\cite{Greilich2011} and \cite{Godden2012}.

The sample consists of a vertically stacked pair of self-assembled InGaAs QDs separated by a 10 nm thick GaAs spacer and embedded within the intrinsic region of a GaAs Schottky photodiode.\cite{Krenner2005} Such devices facilitate complementary photocurrent (PC) and photoluminescence (PL) measurements by varying the internal electric field. For the specific molecule investigated in this paper detailed characterization using PL and PC have been reported in refs.\cite{Mueller11} and \cite{Mueller2012}, respectively. The typical variation of the PL emission with the applied electric field $F$ is presented in the leftmost panel of Fig.1a. We note that all data presented in the manuscript were obtained at $T=10 \, K$. A pronounced anticrossing is seen for electric fields of $F_1 = 23.1 \, kV/cm$ with a splitting of $2V_0 = 3.4 meV$.  This anticrossing arises from tunnel coupling of two excitonic states where the hole is located in the upper QD and the electron resides either in the lowest energy orbital of the upper ($e_{ud}$) or lower ($e_{ld}$) dot, respectively (c.f. insets of Fig. 1(a)). As $F$ increases, the PL intensity quenches due to tunneling of the charge carriers out of the molecule such that PC measurements can be performed. As shown in the rightmost panel of Fig. 1(a), the PC experiments reveal a second anticrossing with a smaller splitting of $2V_1 = 0.8meV$. This feature arises from the coupling of $e_{ud}$ to an excited electron orbital state $e_{ld}^*$ in the lower dot. 

In order to investigate the spin dynamics of charge carriers in the system we performed polarization resolved ultrafast pump-probe experiments with PC readout. To do this, two independently tunable transform limited laser pulse trains with a 3 ps pulse duration were derived from a broadband femtosecond Ti:Sapphire source. \cite{Ruppert2008} While one of them was used to excite the molecule, the photocurrent $I$ induced by the time-delayed probe pulse was measured using a lock-in amplifier.\cite{Zecherle2010} Blocking and unblocking the pump beam revealed the pump induced change of the probe induced photocurrent, $\Delta I$. The quantities $I$ and $\Delta I$ can be interpreted as the linear absorption of the molecules and its \textit{pump induced change}, respectively. To achieve spin selectivity, the optical polarization of the two pulses was adjusted to be either co-circular or cross-circular.

\begin{figure}
\includegraphics[width=1\columnwidth]{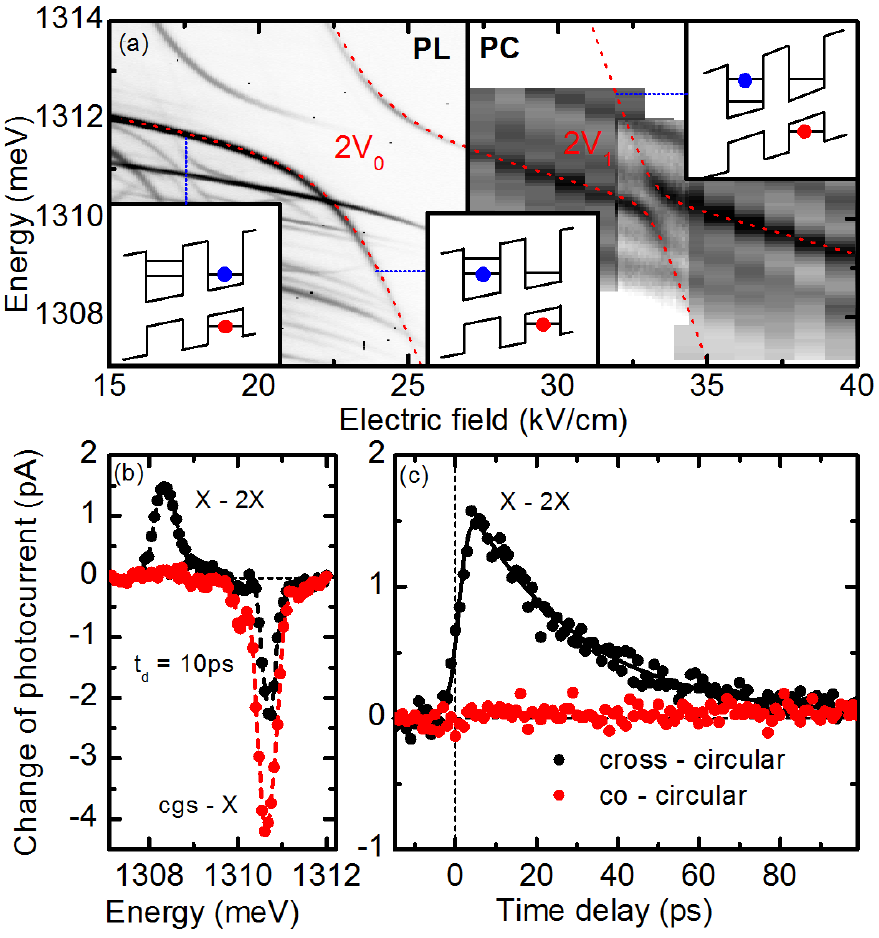}
\caption{\label{fig:Figure_1}
(Color online) (a) Photoluminescence emission (left panel) and photocurrent absorption spectra recorded at T = 10K from the molecule. Anticrossings related to tunnel coupling of the electron in the ground state of the upper dot and the ground (or excited) state of the lower dot as depicted in the insets.(b) Change of PC spectra at $F=30.4 \, kV/cm$ for cross- (co-) circular pump and probe in black (red). (c)Temporal evolution of $\Delta I$ for probing the $X \rightarrow 2X$ transition from (b).}
\end{figure}

To visualize the concept of spin-resolved optoelectronic readout, Fig. 1(b) shows spectra of $\Delta I$ recorded at $F=30.4 \, kV/cm$ with the pump pulse fixed to the transition between the crystal ground state and the neutral exciton in the upper QD at $1310.6 \, meV$ ($cgs \rightarrow X$) and the \textit{tunable} probe pulse, delayed by $10 \, ps$.  Spectra recorded with co- and cross-circular pump and probe pulses are presented by the red and black symbols, respectively. For cross-circularly polarized pulses, we observe a pump-induced bleaching at the $cgs \rightarrow X$ transition energy and pronounced \textit{induced absorption} (at $1308.3 \, meV$) red-shifted by $2.3 \, meV$ from the exciton resonance. The latter transition corresponds to the conditional absorption of the biexciton \cite{Zecherle2010,Mueller2012}. This biexciton absorption is completely absent for the co-circular configuration, an observation attributed to Pauli spin blockade. In addition, the bleaching of the $cgs \rightarrow X$ transition was found to be twice as strong in the \textit{co}-circular configuration. This is mainly a consequence of the present photocurrent measurement that is referenced to the probe beam; while for cross-circular polarizations negative $\Delta I$ results from the bleaching of the $cgs \rightarrow X$ transition only, a co-circularly polarized probe pulse may cause stimulated emission of $X \rightarrow cgs$ giving rise to a further reduction of $\Delta I$. To study carrier spin dynamics, we measured $\Delta I$ at the $X \rightarrow 2X$ transition as a function of the time delay between pump and probe pulse. The results are presented in Fig. 1(c) for cross-circular (black circles) and co-circular (red circles) configurations, respectively. For cross-circular polarizations we observe an exponential decay of the PC change reflecting tunneling of the electron out of the molecule with a time constant of $t_e = 28.4 \pm 1 \, ps$.   This finding is fully consistent with previous results and theoretical simulations.\cite{Ramsay08, Mueller2012,Zecherle2010} In contrast, the co-circular configuration results in no detectable conditional absorption within the electron lifetime, a near perfect Pauli blockade that shows that the spin states are fully robust over the time periods investigated.

\begin{figure}
\includegraphics[width=1\columnwidth]{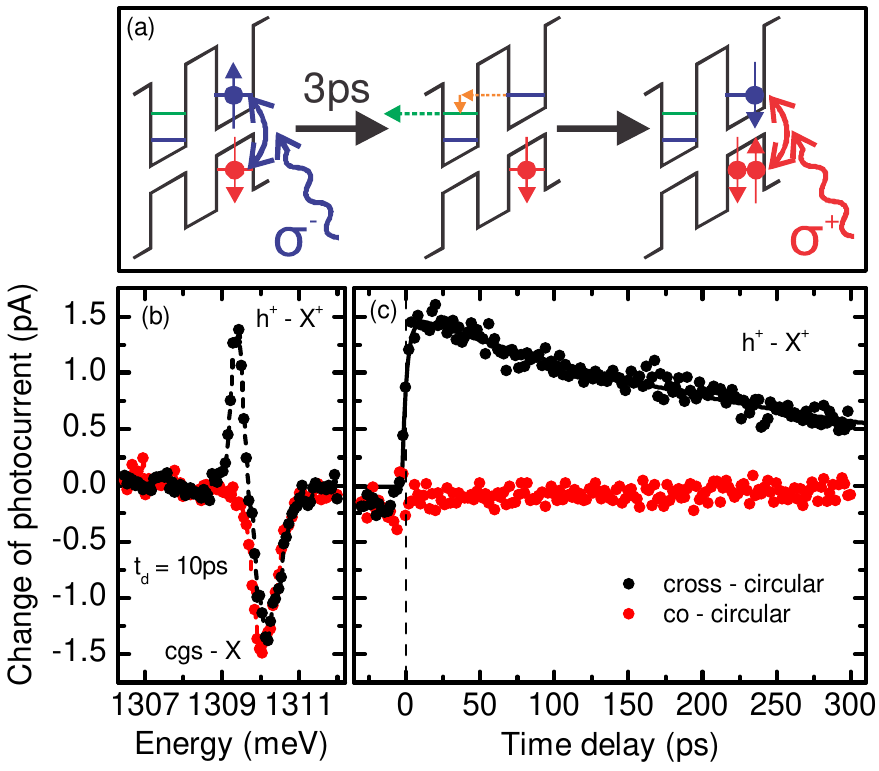}
\caption{\label{fig:Figure_2}
(Color online) (a) Scheme of hole spin creation and read-out. (b) Change of PC spectra at $F=34.1 \, kV/cm$ for cross (co) circular pump and probe in black (red). (c) Temporal evolution of $\Delta I$ for probing the $h^+ \rightarrow X^+$ transition from (b).}
\end{figure}

In the next step, we tune the electric field $F= 34.1 \, kV/cm$ such that an excited orbital state of the electron in the lower dot $e_{ld}^*$ (green level in Fig. 2a) is detuned by $-1.7 \pm 0.1 meV$ lower in energy than the lowest energy orbital state in the upper dot $e_{ud}$ (blue). This configuration is known to result in an ultrafast extraction of the electron out of the upper dot with tunneling times faster than $\sim 4 \, ps$ \cite{Mueller2012}. As depicted schematically in Fig. 2(a), this mechanism relies on inelastic phonon mediated tunneling from $e_{ud}$ to $e_{ld}^*$ which is most efficient for this specific detuning. The intermediate level $e_{ld}^*$ is an excited orbital state and has a small remaining tunnel barrier and thus tunneling to out of the QDM occurs quickly. Spectra of $\Delta I$ for $F= 34.1 \, kV/cm$ and a fixed time delay of $t_D = 10 ps$ after pumping the $cgs \rightarrow X$ transition are presented in Fig. 2(b) as black (red) circles for cross- (co-) circular polarization configurations, respectively. In contrast to the situation in Fig. 1(b), the excitation induced bleaching of the $cgs \rightarrow X$ transition is now found to be similar for both polarization configurations.  This observation reflects the fact that after $t_D = 10 ps$ the electron has already tunneled out of the QD and stimulated emission of $X \rightarrow cgs$ cannot, therefore, occur.  Thus, a negative $\Delta I$ results only from bleaching of the $cgs \rightarrow X$ transition. No absorption $X \rightarrow 2X$ is detected in this data, fully consistent with our expectation of fast electron tunneling. The pump-induced absorption at $1309.4 \, meV$ is detuned by only $0.7 \, meV$ from $cgs \rightarrow X$ arising from the conditional absorption of the $h^+ \rightarrow X^+$ transition of the dot occupied with a single hole that remains after ultrafast electron extraction.\cite{Mueller2012} Most importantly, the $h^+ \rightarrow X^+$ absorption is only present for the cross-circular optical polarizations demonstrating a near perfect spin blockade of the $h^+ \rightarrow X^+$ transition for a co-polarized pump. Consequently, the hole spin is shown to remain practically unchanged during excitation and electron extraction. To investigate this spin blockade in more detail, $\Delta I$ was recorded as a function of the time delay between pump and probe pulse. For the data presented in Fig. 2(c) the pump pulse drives the $cgs \rightarrow X$ transition while the probe pulse is in resonance with the $h^+ \rightarrow X^+$ transition. The cross-circular configuration (black) results in a sharp increase of $\Delta I$ governed by the pulse durations and the short tunneling time $t_e$ of the electron out of the QD. This rise is followed by a much slower exponential decay reflecting the tunnel escape of the hole with a time constant $t_h$. The result of a fit with a rate equation model of sequential electron and hole tunneling is presented as a black line and indicates values of $t_e = 3.1 \pm 1 \, ps$ and $t_h = 308.3 \pm 6.9 \, ps$. In strong contrast to the cross-circular configuration, co-circular polarizations show practically no pump-induced absorption proving a close-to-perfect spin blockade for the entire lifetime $t_h$. In particular, the purity of the spin can be estimated to be $>96 \%$ - limited only by the $150 \, fA$ current noise in measuring $\Delta I$. It is instructive to compare the present results to recent single dot experiments\cite{Ramsay08,Godden2010,Godden2012} where a much less pronounced asymmetry of $t_e$ and $t_h$ is found. In such a situation, the hole spin purity is markedly reduced due to electron-hole exchange interaction while both carriers are present in the dot. Typical values for the purity of the initialization of a hole spin in a single dot using circularly polarized light are $80 \%$ \cite{Ramsay08}. This value might be further enhanced by applying a vertical magnetic field \cite{Godden2012} but significantly decreases to $<50 \%$ in an in-plane magnetic field. Such low purities comprise a severe drawback for coherent control experiments \cite{Godden2012} but could be easily overcome using approaches similar to those developed in our study.

\begin{figure}
\includegraphics[width=1\columnwidth]{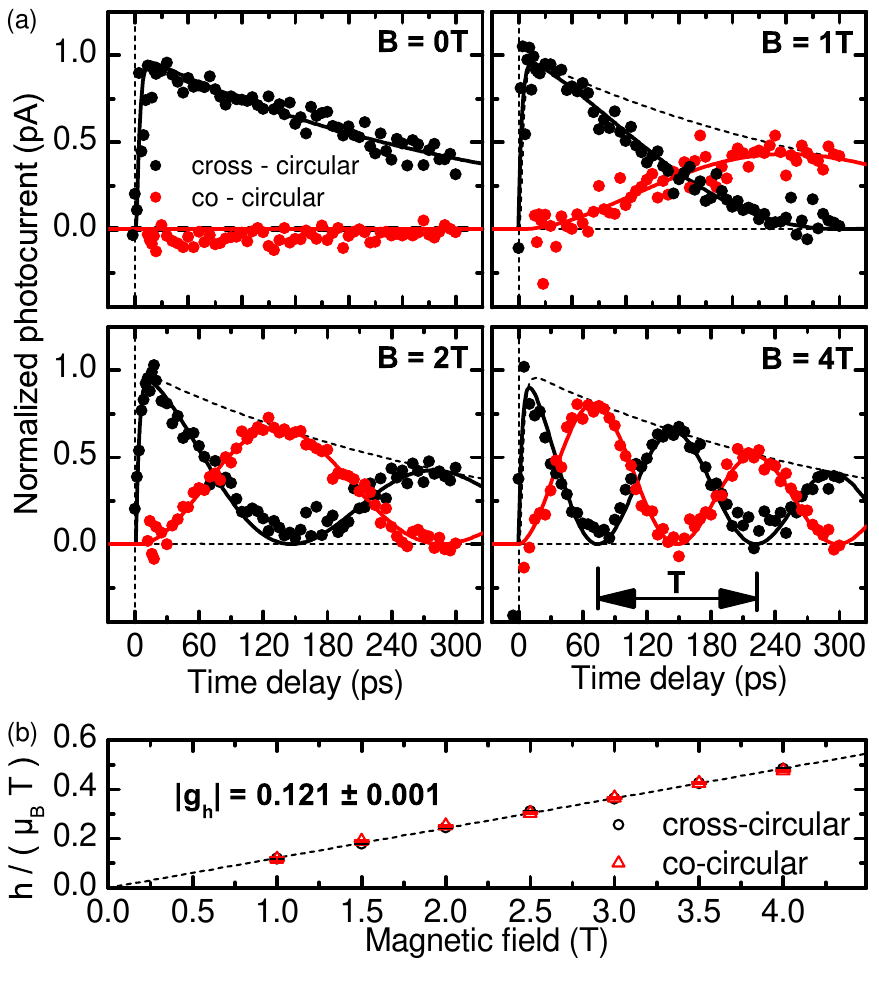}
\caption{\label{fig:Figure_3}
(Color online) (a) Lamor-precession of a single hole spin in an in-plane magnetic field for $B = 0-4 \, T$. Change of PC for cross- (co-) circular pump and probe presented in black (red) reflects the probability to find the hole in the spin up (down) state. Fits presented as solid lines. (b) Analysis of the oscillation periods from (a).}
\end{figure}

As a final step to probe the coherence of the optically prepared hole spin state, we performed magneto-optical experiments with a magnetic field $B$ up to 4 T applied in transverse to the axis of the QD-molecule, perpendicular to the optical axis of our experiment. Selected examples of $\Delta I$ transients are presented in Fig. 3(a) for the probe pulse tuned to the $h^+ \rightarrow X^+$ transition after photogeneration of an exciton. Cross- and co-circular polarization configurations are again represented by the black and red curves, respectively. While for $B=0 \, T$ induced absorptions are only seen for the cross-circular case, this picture changes for non-zero magnetic fields. Most importantly, $\Delta I$ for cross- and co-circular polarizations show pronounced anti-phased oscillations with a frequency that increases with $B$. Since the $\sigma^-$ pump pulse effectively creates a spin up hole the absorption of the probe pulse in the cross- (co-) circular configuration monitors the probability $P_+$ ($P_-$) of the hole being in the up (down) orientation. As a consequence, the oscillation directly monitors the Larmor precession of the single hole spin in the in-plane magnetic field. We used a rate equation model incorporating sequential tunneling of the electron and hole as well as the Larmor precession of the hole spin according to
  
\begin{align}
P_+ = A_+ \left[exp{(-t/t_e)}- exp{(-t/t_h)}\right]\times cos^2(\omega_L t /2)\\
P_- = A_- \left[exp{(-t/t_e)}- exp{(-t/t_h)}\right]\times sin^2(\omega_L t /2)
\label{fits}
\end{align}

with the Lamor-frequency $\omega_L = g_{\bot}\mu_B B /\hbar$ and the Bohr magneton $\mu_B$ to provide quantitative fits to our data. Example fits are presented as solid lines in Fig 3(a), the dashed lines showing the data recorded at zero field, determined primarily by the tunneling of the hole out of the molecule (compare Fig. 2(c)). Most strikingly, the oscillations of the $\Delta I$ transients detected for cross- and co-circular configurations are seen to be fully modulated between the upper envelope and zero. In particular, the fact that $\Delta I$ repeatedly approaches zero for an initially allowed transitions demonstrates negligible loss of coherence during the lifetime of the hole, limited by tunneling. In addition, the magnitude of $\Delta I$ ($A_{\pm} = 1.5 \pm 0.15 \, pA$) and the contrast between the polarization configurations remains the same for all magnetic fields, demonstrating that the purity of the hole spin preparation remains high even at larger magnetic fields. This finding is in strong contrast to single dot experiments where the purity of the hole spin initialization decreases significantly to $<50 \%$ for elevated B \cite{Godden2012}, most likely due to the simultaneous precession of electron and hole spins before the electron tunnels out. However, in our experiment the ultrafast extraction of the electron suppresses interactions between electron and hole so that purity of the initialization remains close to unity. 

In order to quantitatively extract the in-plane hole g-factor we plot $h/ (\mu_B T)$ versus B, where $T$ is the oscillation period of the hole spin precession (cf. Fig. 3(c)). The values of $T$ are obtained from the fits with eq. 1 and 2 to measurements as presented in Fig. 3(a). The values from fitting the cross- (co-) circular transients are depicted in black (red). Both values are in excellent agreement and show a linear B-dependence consistent with $g_{\bot} = 0.121 \pm 0.001$. This measurement is fully consistent with refs. \cite{Greilich2011} and \cite{Godden2012}, indicative of weak hh-lh coupling\cite{Marie1999} and indicating that the coupling to nuclear spin may indeed have near ideal Ising form.\cite{Fischer2008}

In summary we have investigated the dynamics of a single hole spin in a QDM. Perfect Pauli spin blockade in the $X \rightarrow 2X$ and $h^+ \rightarrow X^+$ is observed. Most strikingly, we have realized the initialization of a single hole spin within $3 \, ps$ and with a purity exceeding $96 \%$. In addition, we have measured the Larmor precession of a single hole spin in a lateral magnetic field and demonstrated that the purity of the initialization remains unaffected for magnetic fields up to $4 T$.

We thank A. J. Ramsay for useful discussions and technical advice. We gratefully acknowledge financial support of the DFG via SFB-631, Nanosystems Initiative Munich and the Emmy Noether Program (H.J.K.), the EU via SOLID and the TUM Graduate School. G.A. thanks the TUM Institute for Advanced Study for support.

\end{document}